# Orbital-Selective High-Temperature Cooper Pairing Developed in the Two-Dimensional Limit

Chaofei Liu[1], Andreas Kreisel[2], Shan Zhong[1], Yu Li[1], Brian M. Andersen[3], P. J. Hirschfeld[4], Jian Wang[1,5,6*]

[1]International Center for Quantum Materials, School of Physics, Peking University, Beijing 100871, China
[2]Institut für Theoretische Physik, Universität Leipzig, D-04103 Leipzig, Germany
[3]Niels Bohr Institute, University of Copenhagen, Jagtvej 128, DK-2200 Copenhagen, Denmark
[4]Department of Physics, University of Florida, Gainesville, Florida 32611, USA
[5]CAS Center for Excellence in Topological Quantum Computation, University of Chinese Academy of Sciences, Beijing 100190, China
[6]Beijing Academy of Quantum Information Sciences, Beijing 100193, China

[*]Corresponding author. E-mail: jianwangphysics@pku.edu.cn (J.W.).

**The orbital multiplicity in multiband superconductors yields orbital differentiation in normal-state properties, and can lead to orbital-selective spin-fluctuation Cooper pairing. This phenomenon has become increasingly pivotal in clarifying the pairing 'enigma' particularly for multiband high-temperature superconductors. In one-unit-cell (1-UC) FeSe/SrTiO$_3$, the thinnest and highest-$T_c$ member of iron-based superconductors, the standard electron–hole Fermi pocket nesting scenario is apparently not applicable since the Γ-centered hole pockets are absent, so the actual pairing mechanism is the subject of intense debate. Here, by measuring high-resolution Bogoliubov quasiparticle interference, we report observations of highly anisotropic magnetic Cooper pairing in 1-UC FeSe. From a theoretical point of view, it is important to incorporate effects of electronic correlations within a spin-fluctuation pairing calculation, where the $d_{xy}$ orbital becomes coherence-suppressed. The resulting pairing gap is compatible with the experimental findings, which suggests that high-$T_c$ Cooper pairing with orbital selectivity applies to 1-UC FeSe. Our findings imply the general existence of orbital selectivity in iron-based superconductors and the universal importance of electron correlations in high-$T_c$ superconductors.**

**One-Sentence Summary: Spin-fluctuations-mediated anisotropic Cooper pairing at 2D limit is probed preferentially driven by $d_{xz}/d_{yz}$ orbitals.**

## INTRODUCTION

With a parent Mott-insulating phase, strongly correlated cuprates display low-energy physics dominated by the single Cu-$d_{x^2-y^2}$ orbital. In comparison, iron-based superconductors with moderate correlations possess orbital multiplicity, offering a new opportunity to host rich physics. Within the multiorbital Hubbard model, predominantly depending on Hund's rule coupling $J_H$ over interorbital Coulomb repulsion $U$ (*1*), the so-called 'Hund's metal' state can emerge (*2*). Essentially, the Hund's metal is a special metallic state 'fingerprinted' with orbital differentiation due to a suppression of interorbital charge fluctuations (*3-5*). This orbital-decoupling effect switches the multiband system from collective to mutually independent, individual orbital behavior. In consequence, the physics strongly depends on the individual filling and electronic structures of each orbital separately (*6*). For the Fe atom, the statistically different fillings for five 3$d$ orbitals occupied by six electrons correspond to different degrees of proximity to the half-filling Mott insulator (*6, 7*), consistent with the moderately correlated, bad-metallic nature of multiband superconductors. Naturally, Cooper pairing can also be orbital-selective and accordingly anisotropic (*8*), where electrons of specific orbital(s) primarily bind to form the Cooper pairs in certain momentum directions.

Among the family of iron-based superconductors, the 11 iron chalcogenides are relatively strongly correlated (*2, 9*), compared to the 111, 122, and 1111 systems. Therefore, FeSe provides a desirable platform for studying orbital-selective Cooper pairing. Previously, the experimental discovery of orbitally selective pairing was limited to bulk FeSe (*10*), with its counterpart in the two-dimensional (2D) limit, i.e., one-unit-cell (1-UC) FeSe (*11, 12*), scarcely explored. In particular, for 1-UC FeSe film on SrTiO$_3$(001), although the significantly high $T_c$ of 55–65 K has encouraged extensive investigations (*13*), the central pairing issues are still under debate (*14-17*). Due to the absence of hole pockets at the Brillouin-zone (BZ) center, the generally accepted $s_\pm$-wave pairing structure based on electron–hole Fermi nesting is conceptually inapplicable. Alternative pairing conjectures were proposed theoretically, mainly including $s_{++}$-, incipient $s_{\pm}$-, extended $s_{\pm}$-, and nodeless $d$-wave (9, 15). A consensual pairing understanding is still lacking, in part due to the experimental challenge of accurately distinguishing different types of bosonic modes (e.g., spin fluctuations vs. phonon) for mediating the coherent Cooper pairs. Revealing orbital-selective pairing would be crucial to clarifying the widely debated pairing mechanism for 1-UC FeSe. However, whether orbital-selective Cooper pairing can survive in FeSe in the 2D limit is not clear. There are several considerations: i) In the Hund's metal states of iron selenides, the localized $d_{xy}$ orbital is more correlated than the itinerant $d_{xz}/d_{yz}$ orbitals (18-21); however, in 1-UC FeSe, the forbidding of $d_{xz}/d_{yz}$ interlayer hoppings between adjacent Fe–Se layers tends to localize the $d_{xz}/d_{yz}$ orbitals more, suppressing their orbital differentiation with $d_{xy}$ regarding correlation strength. ii) Orbital selectivity decreases with electron concentration for Fe $3d$ orbitals (6), and thus might be expected to be weak in heavily electron-doped 1-UC FeSe. iii) A finite nematic order serves to stabilize the selective Mottness (22); hence, the absence of nematicity in 1-UC FeSe may also diminish orbital selectivity.

## RESULTS AND DISCUSSIONS

### Superconducting gap anisotropy

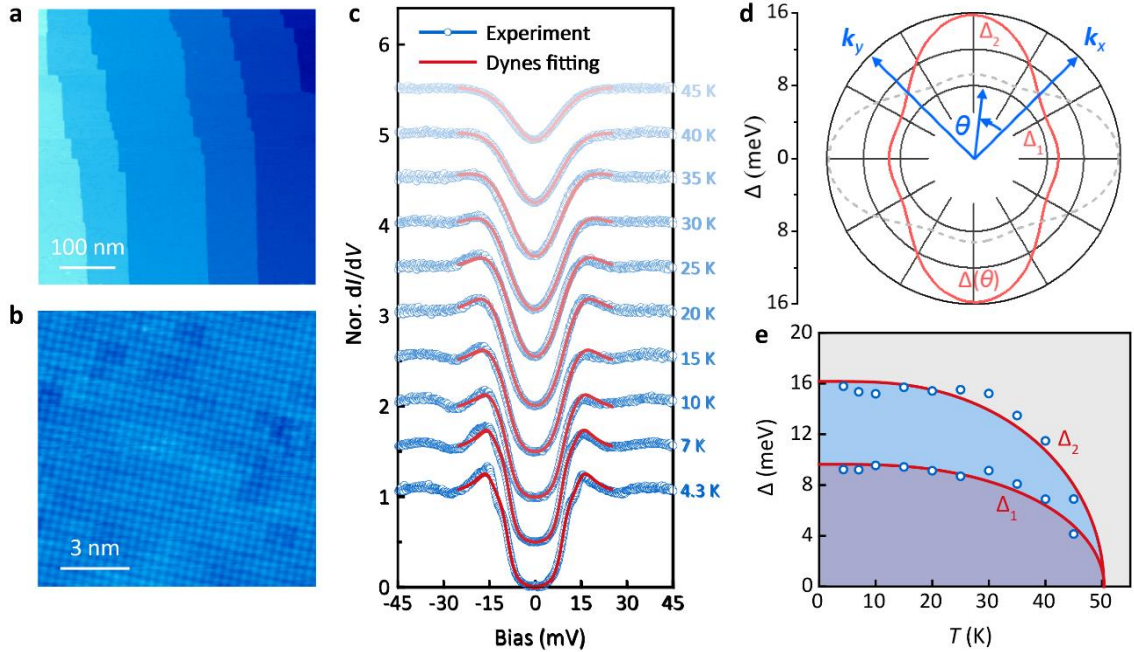

**FIG. 1. SC gap anisotropy in crystalline 1-UC FeSe/SrTiO$_3$.** (a,b) Topographic images of 1-UC FeSe/SrTiO$_3$. Size: (a) 500×500 nm$^2$, (b) 12×12 nm$^2$; set point: (a) $V$ = 1 V, $I$ = 500 pA, (b) $V$ = 0.9 V, $I$ = 500 pA. (c) Temperature dependence of the normalized experimental tunneling spectra (open symbols) (vertically offset for clarity), which are fitted by anisotropic Dynes function (solid curves) within the bosonic-coupling-uninfluenced bias window ~[−25,25] mV. All the spectra shown throughout are taken at defect-free regions to avoid the influence of defect-induced bound states. Set point: $V$ = 0.04 V, $I$ = 2500 pA; modulation: $V_{mod}$ = 0.5 mV (by default). (d) Examples of SC gaps $\Delta(\vartheta)$ (solid line) on unhybridized electron pocket used for the Dynes fittings (4.3 K) in (c). Here, $\vartheta$ is defined relative to $k_x$ (anticlockwise, $+\vartheta$) in the folded BZ, which is along the gap-minimum direction (24). $\Delta_{1,2}$, $\Delta(\vartheta)$ maxima ($\Delta_1 < \Delta_2$). $\Delta(\vartheta)$ obtained by BZ folding (dashed line) is shown for comparison. (e) BCS fittings (solid curves) to the temperature-dependent $\Delta_{1,2}(T)$ (open symbols) obtained from the Dynes fittings in (c). For Dynes and BCS fitting details, see Supplementary Section II.

Here, by spectroscopic-imaging scanning tunneling microscopy (STM) (4.3 K unless specified), we detected in 1-UC FeSe a highly anisotropic magnetic Cooper pairing preferentially driven by $d_{xz}/d_{yz}$ orbitals. The 1-UC FeSe film was epitaxially grown on Nb:SrTiO$_3$(001) following the well-developed recipe (23). Topographic imaging reveals *in situ* the high crystalline quality at both mesoscopic and microscopic scales [Figs. 1(a) and 1(b); for more STM characterizations, see Supplementary Section I]. With increasing temperature $T$, the tunneling spectrum (d$I$/d$V$ vs. $V$) is thermally smeared as expected, showing multigap-type coherence peaks broadened progressively [Fig. 1(c)]. Within the bosonic-coupling-unmodified bias region [−25,25] mV, the $T$-dependent spectra were fitted by Dynes function (Supplementary Section II). Strikingly, the anisotropic Dynes formula shows reasonable fittings with appropriate pairing strength $\Delta_{1,2}$ ($\Delta_1 < \Delta_2$) and thermal broadening $\Gamma$ [Fig. 1(c)]. The involved angular-dependent $\Delta(\theta)$ are shown in Fig. 1(d) for the 4.3-K spectrum, exhibiting alternating gap maximum and minimum as observed in high-resolution angle-resolved photoemission spectroscopy (ARPES) experiments (*24*). As for the Dynes fittings without incorporating the anisotropic superconducting (SC) gaps, the multigap-type spectra of 1-UC FeSe evidently cannot be fitted by the isotropic Dynes formula that only gives single pair of coherence peaks. For two summed Dynes functions with different weights, while part of the tunneling spectra can be fitted, the parameters, especially $\Delta_{1,2}$ and $\Gamma_{1,2}$, are physically unreasonable for all five measured sets of temperature-dependent spectra (Supplementary Section II). Moreover, the obtained $\Delta_{1,2}(T)$ from anisotropic Dynes fittings can be fitted by the Bardeen–Cooper–Schrieffer (BCS) gap equation [Fig. 1(e)], giving moderate transition temperatures $T_{c1,2}$ (~50 K) compared with those (~60–65 K) empirically extrapolated from zero-bias differential conductance at different temperatures (Supplementary Section III). As already mentioned, the form of the gap used to obtain the anisotropic fits is as shown in Fig. 1(d). Therefore, the 'successful' fittings to the tunneling spectra selectively by ARPES-consistent anisotropic Dynes function indirectly show the signature of SC-gap anisotropy by an *in situ* tunneling-spectroscopic technique.

## High-resolution imaging of anisotropic quasiparticle interference

The electronic structures of a material can be reflected in the energy-dependent Bogoliubov quasiparticle-interference (BQPI) patterns. For 1-UC FeSe, quasiparticle scattering consists of intra- and inter-pocket components, $\boldsymbol{q}_1$, $\boldsymbol{q}_2$ and $\boldsymbol{q}_3$ [$\boldsymbol{q}$, momentum ($\boldsymbol{k}$) transfer] [Fig. 2(c)], which are symmetry-equivalent (*25*) since only differing by a reciprocal lattice vector, and are here defined in the *folded* BZ for the 2-Fe unit cell [e.g., see Fig. 4(a); adopted *throughout* unless otherwise specified]. Furthermore, due to tunneling-matrix-element effects, the scattering intensities of $\boldsymbol{q}_1$, $\boldsymbol{q}_2$ and $\boldsymbol{q}_3$ decrease successively (*25*). The symmetry equivalence and intensity difference of $\boldsymbol{q}_1$–$\boldsymbol{q}_3$ motivate us to mainly focus on the highest-intensity $\boldsymbol{q}_1$ scattering in the following. In general, the resolution of BQPI patterns can be controlled by the atomic configurations of STM tips. Typically, lower-spatial-resolution tips are more sensitive to long-wavelength modulations, corresponding to low-$\boldsymbol{q}$ scatterings (*10, 26*). To image the structure of $\boldsymbol{q}_1$ scattering with relatively higher resolution, we thus intentionally decorated the tip until a high resolution for low-$\boldsymbol{q}$ scatterings is achieved, at the expense of high-$\boldsymbol{q}$ resolution (Supplementary Section V).

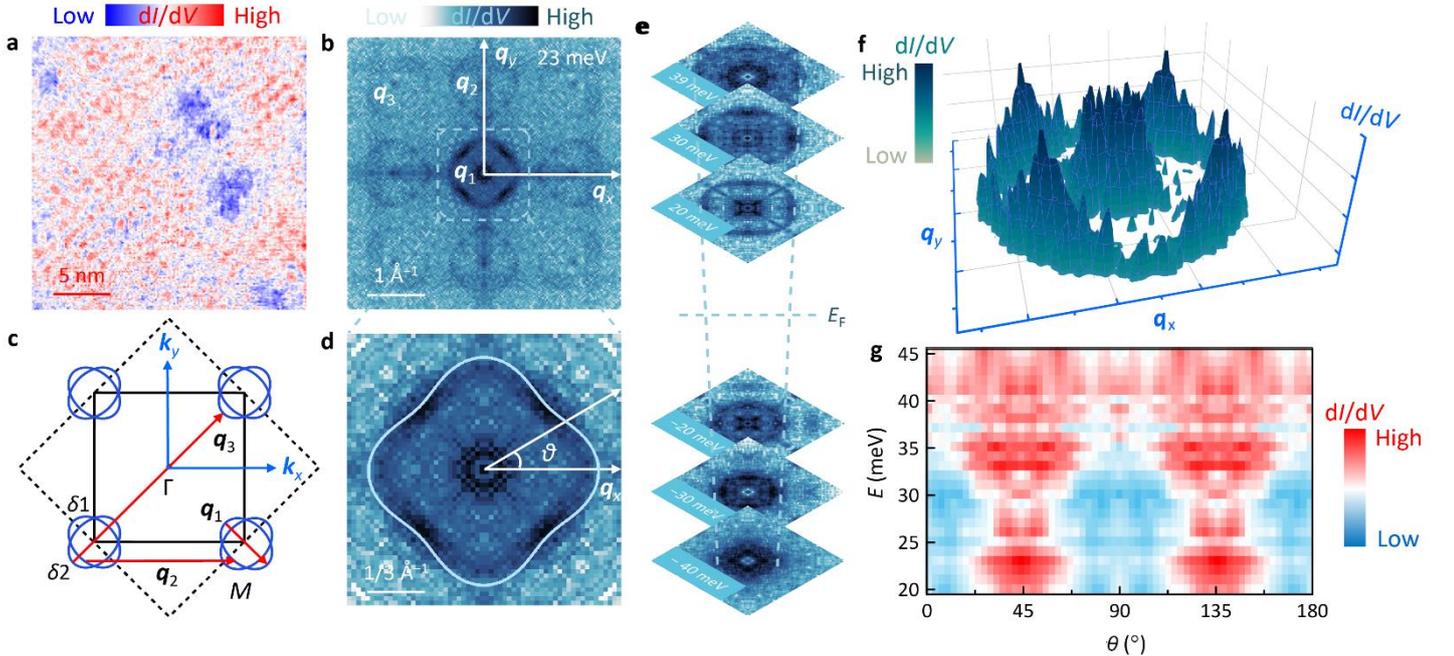

**FIG. 2. High-resolution imaging of the anisotropic BQPI.** (a,b) Selected BQPI d$I$/d$V$($\boldsymbol{r}$,$E$) mapping (24×24 nm$^2$) and corresponding FFT pattern at energy $E$ = 23 meV. For BQPI-processing details and more BQPI patterns, see Supplementary Section IV. (c) Fermi-surface topology of 1-UC FeSe/SrTiO$_3$ in the folded BZ (solid square). Here, spin–orbit coupling, which would induce a small $\delta_1$/$\delta_2$-pocket hybridization, is ignored. $\boldsymbol{q}_1$, $\boldsymbol{q}_2$ and $\boldsymbol{q}_3$ schematically denote the classes of possible intra- and inter-pocket scattering vectors. Their resulting scattering pockets are accordingly assigned in (b) also as $\boldsymbol{q}_1$–$\boldsymbol{q}_3$, where $q_x$/$q_y$ are defined based on $q$–$k$ space correspondence. (d,f) 2D and 3D plots of the zoom-in view of $q_1$ pocket in (b). The closed line in (d) [$q = 0.576 + 0.0654\cos 4\vartheta$ (Å$^{-1}$)] sketches the profile of $q_1$ pocket. To guarantee the angular-definition consistency with Fig. 1(d), $\vartheta$ here is defined relative to the $q_x$ (anticlockwise, +$\vartheta$). (e) Stacked FFT-BQPI $q_1$ patterns at typical $E$ across the Fermi energy $E_F$. (g) $\vartheta$-dependent FFT-d$I$/d$V$($\vartheta$,$E$) along the $q_1$-pocket profile as depicted in (d) at different $E$.

In the fast-Fourier-transformed (FFT) BQPI patterns for obtained d$I$/d$V$(**r**,$E$) mappings at positive energies $E$ [e.g., Figs. 2(a) and 2(b)], the $q_1$ scattering pockets are selectively high-resolved. This is in contrast to previous reports (*27*), where the $q_1$ pockets are smeared out by the low-$q$ scatterings caused by spatially random defects. As the energy decreases across the Fermi level $E_F$, the size of $q_1$ pocket shrinks accordingly [Fig. 2(e)], confirming the electron-type nature of the weakly hybridized Fermi pockets $\delta_1$/$\delta_2$ at $M$ points [Fig. 2(c)]. 2D and 3D plots of the zoom-in views for $q_1$ pocket, e.g., at $E = 23$ meV [Figs. 2(d) and 2(f)], intuitively present the fine structures with significant anisotropy. More quantitatively, as shown in Fig. 2(g), angular-dependent FFT-d$I$/d$V$($\theta$,$E$) [$\theta$ consistently defined in Figs. 1(d) and 2(d)] was extracted along the profile of the $q_1$ pocket as approximately described by $q = 0.576 + 0.0654\cos 4\theta$ (Å$^{-1}$) [Fig. 2(d)] at different $E$, signifying the anisotropy directly visualized in FFT-BQPI images. (Notice that the BQPI data reported in Ref. (*16*) for 1-UC FeSe exhibit similar signatures of $q_1$-pocket anisotropy. Yet the analyses therein mostly focus on the energy dependence of azimuthally integrated BQPI intensity of $q_1$–$q_3$. In contrast, as discussed in detail below, our BQPI analyses are instead mainly concentrated on $q_1$ dispersions $E(q_1,\theta)$ and related self-energy effect in different angles benefiting from our highly resolved $q_1$ pattens.) In addition, the periodic changes of FFT-d$I$/d$V$($\theta$,$E$) preferentially scale with $\Delta_{\text{avg.}}(\theta) = [\Delta(\theta)+\Delta(\theta+90°)]/2$ [$\Delta(\theta)$ as defined in Fig. 1(d)], both showing peaks at $\theta = 45°+90°\times N$ ($N = 0, 1, 2, 3$). [In view of the hybridization of $\delta_1$/$\delta_2$ pockets and the lack of separate $\delta_1$/$\delta_2$ resolution in BQPI, $\Delta(\theta)$ on crossed ellipse-like $\delta_1$/$\delta_2$ pockets behaves as the averaging result $\Delta_{\text{avg.}}(\theta)$ of $\Delta(\theta)$ and $\Delta(\theta+90°)$ on unhybridized electron pockets (*28*). Despite the C2-symmetric $\Delta(\theta)$ and $\Delta(\theta+90°)$ on unhybridized pocket, the averaged $\Delta(\theta)$, $\Delta_{\text{avg.}}(\theta)$, is C4-symmetric as the BQPI pattern.]

**Anisotropic self-energy effect**

In the microscopic theory of superconductivity, electrons bind to form Cooper pairs via the exchange of virtual bosons [Fig. 3(a)]. In view of the detected pairing anisotropy, clarifying the nature of the pairing interaction, i.e., electron–boson coupling, is of immediate interest. In strong-coupling Eliashberg theory for incorporating the electron–boson coupling, the noninteracting electronic dispersion $\varepsilon(\boldsymbol{k})$ and single-particle density-of-states (DOS) spectrum $N(E)$ originally described by mean-field theory are modified. These modifications typically result in additional 'kinks' and 'humps' in $\varepsilon(\boldsymbol{k})$ and $N(E)$, respectively, e.g., at $E^\Sigma = \Delta_2+\Omega$ ($\Omega$, boson energy) for a multiband superconductor [Figs. 3(c) and 3(d)]. Previously, the renormalized kink and hump anomalies have been intensively investigated as the signatures of pairing-related electron–boson coupling, especially in high-temperature cuprate superconductors (*26*). In the interacting Green's function theory, the many-body effect likewise is encapsulated in the *complex* self-energy $\Sigma(\boldsymbol{k},E)$ [Fig. 3(b)] for the corresponding noninteracting state $|\boldsymbol{k},E\rangle$. In detail, the real part Re$\Sigma(\boldsymbol{k},E)$ describes the deviation from bare band dispersion $\varepsilon(\boldsymbol{k})$, and the imaginary part Im$\Sigma(\boldsymbol{k},E)$ describes the energy broadening by finite lifetime (*29*). Naturally, the singularity in energy-dependent $\Sigma(\boldsymbol{k},E)$ at $E^\Sigma = \Delta+\Omega$ yields the bosonic kink (or hump) in the electronic spectrum.

For iron-based superconductors, the proposals regarding specific exchanged bosons for mediating Cooper pairing have been controversial, including suggestions of antiferromagnetic spin fluctuations (AFSF) with peaked spin susceptibility (*30*), and $d$-orbital fluctuations induced by electron–phonon interaction (*31*). As discussed in the following, the differentiation of these distinct types of electron–boson coupling is encoded in the self-energy effects with different characteristics (*32*). For 1-UC FeSe, the phonon suggested to participate in enhancing pairing potential is the optical branch with flat dispersion [upper inset of Fig. 3(c)], rather than the highly dispersive acoustic branch (*33*). By contrast, within proposed extended $s_\pm$- and nodeless $d$-wave pairings, the related $\boldsymbol{Q} \approx \langle 2\pi,0\rangle$ AFSF are strongly momentum-dependent [lower inset of Fig. 3(c)]. Here, $\langle 2\pi,0\rangle$ denotes the $(2\pi,0)$-equivalent vectors [$(\pm 2\pi,0)$ and $(0,\pm 2\pi)$] in unit of $1/a_0$ ($a_0$, lattice constant) in the *folded* BZ (i.e. $\langle\pi,\pi\rangle$ in *unfolded* BZ). The self-energy $\Sigma(\boldsymbol{k},E)$, which can describe the electron–boson coupling, accordingly shows the momentum structures highly dependent on the concrete type of involved bosons (*32*). To be specific, for a multiband system coupled to AFSF sharply peaked at momenta ~<2π,0>, the kinematic constraints $E_{\mathbf{k}}^{j} = E_{\mathbf{k-Q}}^{i} - \Omega$ (*i*,*j*, band indexes) require the electron–AFSF coupling, i.e., the self-energy effect, be accordingly momentum-dependent (anisotropic), reminiscent of the above-observed anisotropic pairing. Similarly, a multiband system coupled to a dispersionless phonon is expected to exhibit momentum-independent (isotropic) self-energy. The anisotropic band renormalization due to the self-energy by AFSF coupling is evidently in sharp contrast to the isotropic behavior by phononic coupling, providing a highly promising approach in differentiating AFSF and phonon modes.

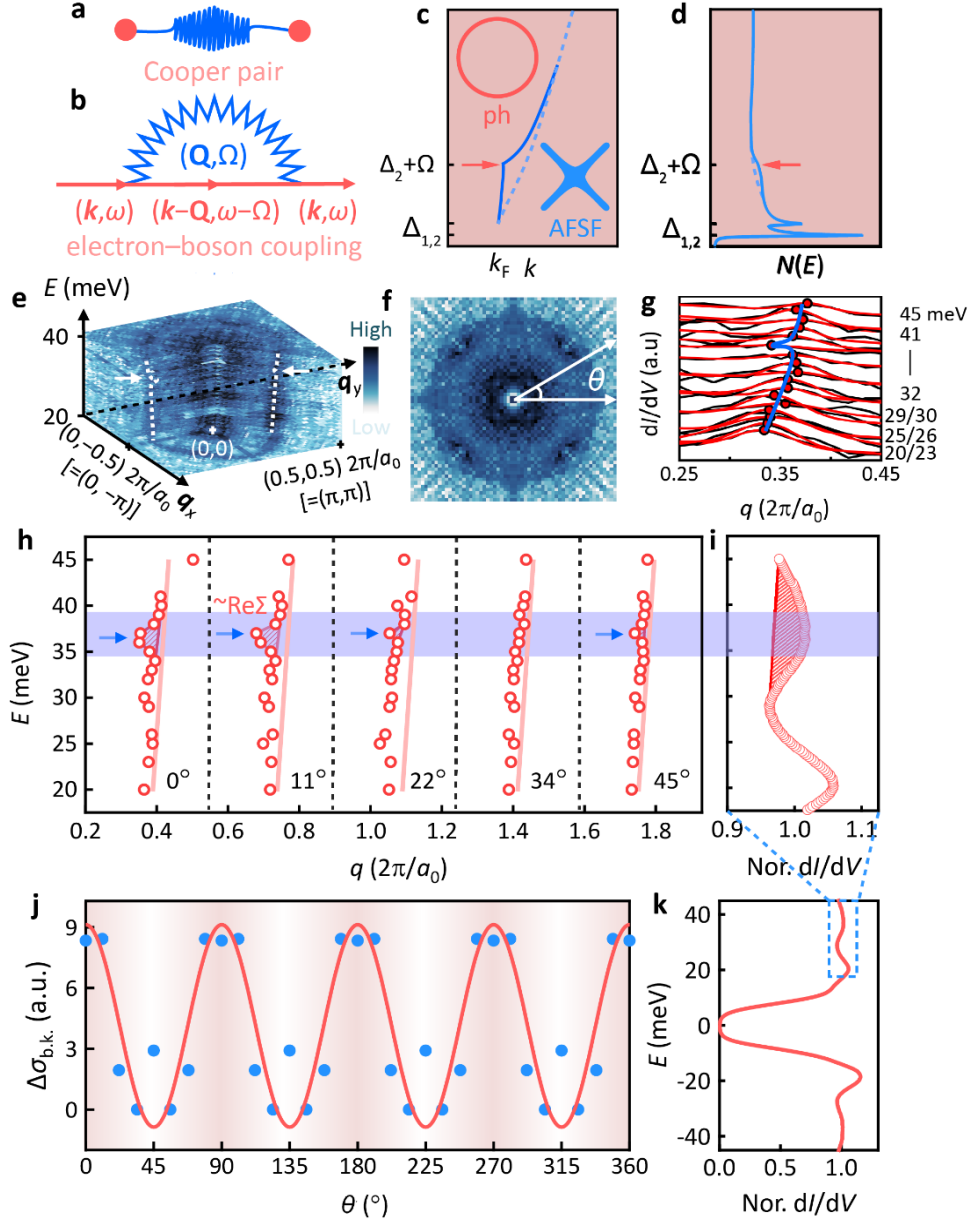

**FIG. 3. Anisotropic self-energy effect in experiments.** (a,b) Cartoon of the boson-mediated Cooper pair and diagram of the lowest-order self-energy induced by electron–boson coupling. (c,d) Electronic dispersion *E*(*k*) and quasiparticle-DOS spectrum *N*(*E*) (reproduced from ref. (*23*)) both modified by electron–boson coupling, resulting in a kink and a hump (arrows), respectively, at $\Delta_2+\Omega$ for the multi-band superconductivity. The insets in (c) are the schematics of the *momentum* distributions of isotropic phonon and anisotropic AFSF in BZ, showing the flat and highly dispersive momentum structures, respectively. (e) FFT-d*I*/d*V*($q_x,q_y,E$) for the $\mathbf{q}_1$ pocket plotted in *E*–*q* space. The kinks (arrows) in $\mathbf{q}_1$ dispersions (dotted lines) are highlighted along $(0,-0.5)2\pi/a_0$ [=(0,−π)] and $(0.5,0.5)2\pi/a_0$ [=(π,π)] directions. (f) Illustrating the $\vartheta$ definition in FFT-BQPI $\mathbf{q}_1$ pattern. (g) Lineplots of FFT-d*I*/d*V*(*q*,*E*) at different *E* (*E* ∈ [20,45] meV) (vertically offset for clarity) for $\vartheta$ = 45° (azimuthally averaged over $\vartheta$±5°), exemplifying the extraction method of the dispersion *E*(*q*) for $\mathbf{q}_1$ pocket. The red lines are the Gaussian fittings with linearly tilted background, whose peak positions (red symbols) give the *q* of FFT-d*I*/d*V*(*q*,*E*)-linecut maxima at different *E*. (h) Extracted *E*(*q*) along representative directions (horizontally offset for clarity). The spectral weights

Δσ$_{b.k.}$ of $E(q)$ modifications (arrows) with respect to the noninteracting backgrounds (solid lines) are shown as hatched areas, which can be used to approximately quantify the real part of self-energy ReΣ($E^Σ$ ~37 meV). (i,k) Typical normalized tunneling spectrum for 1-UC FeSe, where the bosonic hump is highlighted in shadow. (j) Measured Δσ$_{b.k.}$ (solid symbols) as a function of $\vartheta$, overlaid by the eye-guiding line for clarity of the anisotropy.

Experimentally, the self-energy effect can be detected by ARPES by measuring the spectral function $A(k,E)$, yet only for the occupied states below $E_F$. However, in previous experiment (23), the bosonic mode was found coupling more strongly to electron-like states above $E_F$ for 1-UC FeSe (23). Therefore, for the study of self-energy in 1-UC FeSe, STM-based BQPI is preferred because of the capability of accessing the unoccupied states above $E_F$. Figure 3(e) plots the complete representation of measured FFT-BQPI d$I$/d$V(q,E)$ in $E$–$q$ space for the empty states, with (0,−π) and (π,π) directions highlighted. Preliminarily, in this $E$–$q$ presentation, the 'exposed' dispersions $E(q)$ (dotted lines) for the $q_1$ pocket are seemingly not equivalent in different directions [e.g., (0,−π) vs. (π,π)]. [Given the relation $q_1=2(k\pm G')$, where $G'$=<π,π>, $E(q)$ for $q_1$ carries conceptually the information of electronic dispersion $E(k)$.] Inspired by the contrasting self-energy effects for different bosonic modes, we then extracted the angular-dependent $E(q)$ from the linecuts of FFT-BQPI patterns at different energies [Figs. 3(f)–3(h)], which is commonly believed highly challenging for STM technique. The positive slope of obtained $E(q)$ [Fig. 3(h)] agrees quantitatively with the aforementioned electron-type nature of the $M$-centered Fermi pockets [Fig. 2(e)]. More intriguingly, $E(q)$, particularly for $θ=0°$ and 11°, shows band-renormalized kinks at $E^Σ$ ~37 meV, coinciding in energy with the bosonic hump detected in d$I$/d$V$ (i.e., DOS) spectrum [Figs. 3(i) and 3(k)] (23). Strictly speaking, the inelastic tunneling can be additionally considered to give rise to the bosonic hump (34), which is typically ascribed to spin or nematic fluctuations in specific iron-based superconductors (e.g. LiFeAs) (35, 36). In spite of these facts, the kink–hump correspondence at a common energy scale implies the true existence of the reconstructions in $E(q)$ with a bosonic-coupling origin.

In the pure two-band scenario, the bosonic kink is selectively resolved at $Δ_2+Ω$ (band #2) instead of $Δ_1+Ω$ (band #1) (23). For $E(q)$ extracted from the FFT-BQPI data without direct band resolution, the evident kink feature [Fig. 3(h)] is thus actually induced in band #2. To quantify the self-energy effect, we integrated approximately as ReΣ($E^Σ$ ~37 meV) the shadowed spectral weight Δσ$_{b.k.}$ of the bosonic kink deviating from the noninteracting background [Fig. 3(h)] (for $θ=22°$ and 45°, the relatively weak bosonic kinks can be accordingly assigned at $E^Σ$ ~37 meV in view of the pronounced kink anomalies clearly identifiable for $θ=0°$ and 11°; for alternative ReΣ definition, see Supplementary Section VI.A). Δσ$_{b.k.}$($θ$) extracted at different directions is plotted in Fig. 3(j). Remarkably, Δσ$_{b.k.}$ is highly anisotropic with fluctuating amplitude of an intrinsic nature (Supplementary Section VI.B). In particular, Δσ$_{b.k.}$($θ$) peaks at $θ$ = 90°×$N$, consistent with the peaked positions of inter-pocket pair scatterings $Q$ ≈ <2π,0> for AFSF in extended $s_±$- and nodeless $d$-wave pairings for 1-UC FeSe. Based on the above-introduced different self-energy effects induced by AFSF and a phonon mode, the striking anisotropy in Δσ$_{b.k.}$($θ$) directly points to an AFSF-coupling explanation of the bosonic kink, leaving the phonon mechanism likely irrelevant. All above results combined together highlight an anisotropic pairing possibly mediated by ~<2π,0> AFSF coupling (Supplementary Section VI.C) (37).

**Orbital-selective Cooper pairing**

As summarized in Fig. 4(b), the major findings in our study consist of the two-gap properties of the spectra with distinct $Δ_{1,2}(T)$ leading to anisotropic gaps $Δ(θ)$ contributed by the Cooper pairing via $Q$ ≈ <2π,0> AFSF. These AFSF nest the weakly hybridized and sign-reversed electron pockets at adjacent $M$ points, although $Q$ is not a perfect Fermi-nesting vector. Especially, the concluded ~<2π,0> AFSF from anisotropic self-energy can preliminarily exclude $s_{++}$- and incipient $s_±$-wave, constraining the dominated pairing scenario with a high possibility as extended $s_±$- or nodeless $d$-wave. Besides our self-energy analysis based on the higher-energy BQPI data, other methods focusing on the lower-energy BQPI results can alternatively address the pairing state. A common strategy for analyzing these BQPI data near $E_F$ can be deducing the momentum structure of SC gaps, and then comparing with calculations under different pairing scenarios (10).

Yet, the exact physical nature of the SC-gap anisotropy remains unclarified. In 1-UC FeSe, electron correlations are predominantly dependent on Hund's coupling $J_H$ (38), and the orbital-selective physics would be dominant in principle.

Correspondingly, the Cooper pairing is expected to be orbital-selective, where, based on $\Delta(\mathbf{k}) = \sum_i \Delta_i(\mathbf{k}) W_i(\mathbf{k})$ [$W_i(\mathbf{k})$, weight of orbital $i$], electrons of dominant weight primarily participate to form the Cooper pairs. The SC gaps are thus expected to be highly anisotropic because of the orbital anisotropy, with magnitude following the weight of specific orbital(s) over the (unhybridized) Fermi pocket (*8*).

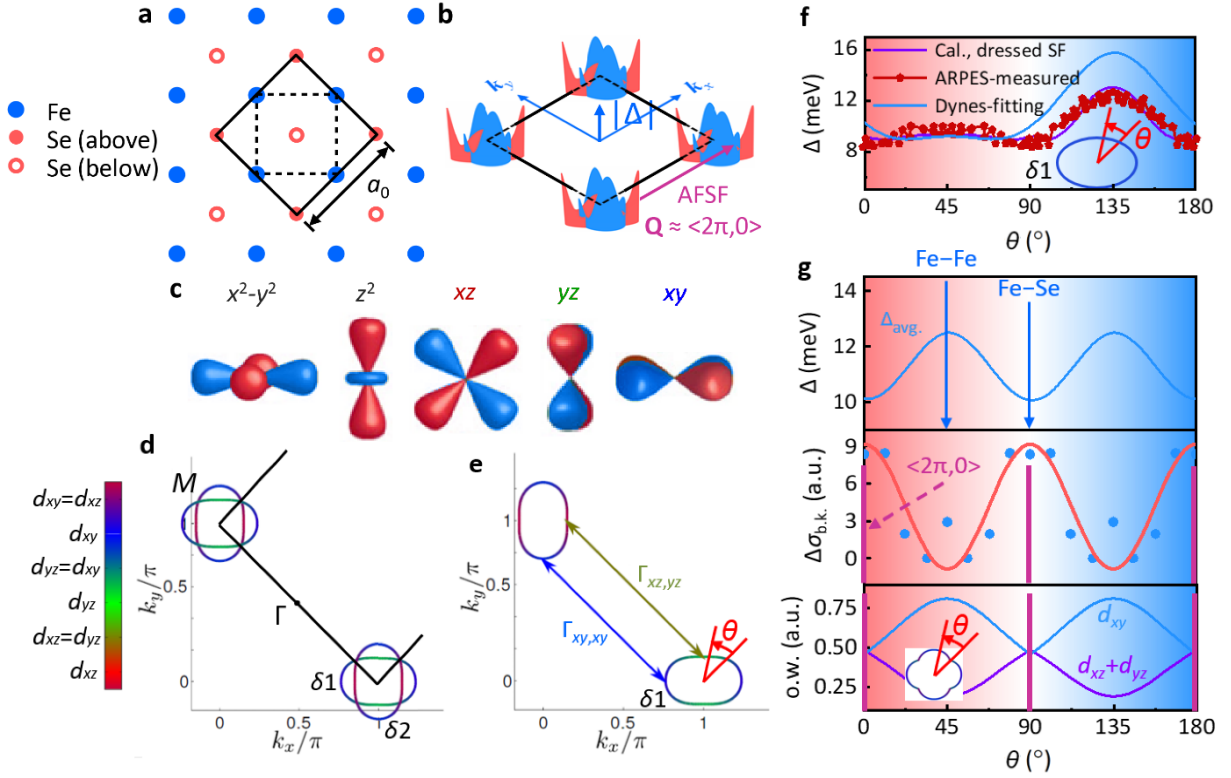

**FIG. 4. Orbital-selective Cooper pairing.** (a) Schematic of the FeSe lattice. The solid square denotes the 2-Fe unit cell that corresponds to the folded BZ. (b) Summary of the measured SC-gap structure in the folded BZ. The sign of the gap function is encoded by different colors (exemplified within nodeless *d*-wave). (c) Cartoons of the Fe 3*d* orbitals. (d) Weight distributions of the Fe 3*d* orbitals over the Fermi surface from our tight-binding parametrization (*30*). (e) Relevant pairing channels as derived from spin-fluctuation pairing mechanism. $\Gamma_{xy,xy}$ ($\Gamma_{xz,yz}$), intra-$d_{xy}$-orbital (inter-orbital, $d_{xz} \leftrightarrow d_{yz}$) pair-scattering strength. (f) Theoretical $\Delta(\vartheta)$ on unhybridized electron pocket $\delta_1$, obtained based on the spin-fluctuation (SF) approach including orbital-dependent quasiparticle weights ($\sqrt{Z_{xy}}$ =0.4273, $\sqrt{Z_{xz}} = \sqrt{Z_{yz}}$ =0.9826) (*30, 42*). The ARPES-measured (*24*) and Dynes-fitting $\Delta(\vartheta)$ are plotted for direct comparison. (g) $\vartheta$ dependences of $\Delta_{\text{avg.}}(\vartheta)$, $\Delta\sigma_{\text{b.k.}}(\vartheta)$, and orbital weights (o.w.) of $d_{xy}$ and $d_{xz}+d_{yz}$ on outer hybridized pockets. $\Delta(\vartheta)$ is plotted as $\Delta_{\text{avg.}}(\vartheta)$ here given the lack of separate $\delta_1/\delta_2$ resolution in folded BZ for BQPI-measured $\Delta\sigma_{\text{b.k.}}(\vartheta)$. The angle positions corresponding to <2π,0> are highlighted by the purple solid lines in the middle and bottom panels.

To determine whether the SC-gap anisotropy is orbital-driven, we present the orbital weights from a tight-binding model for the five Fe 3*d* orbitals [Fig. 4(c)] over the Fermi surface [Fig. 4(d); see Supplementary Section VII for details] (*30*). The Se 4*p* orbitals are neglected because of their negligible DOS near $E_F$, which is several orders of magnitude lower than that of Fe 3*d* orbitals (*39*). A more quantitative theoretical approach using Wannier functions for the surface layer reveals that, tails of the Wannier functions originating from Se surface atoms dominate the real space imaging (*40*). For the analysis of the BQPI at small wave vectors **q**, the modulations at length scales of the unit cell or smaller, such as the intra-unit-cell shapes and features of the Wannier states in which tunnelling occurs, do not play a role. The pairing structure can be understood in terms of orbital-dependent pairing interactions $\Gamma_{a,b}$ such that the intra-orbital, inter-pocket pairing interaction $\Gamma_{xy,xy}$ via the weight-dominated $d_{xy}$ channel naturally drives the gap maxima at the pocket 'convex ends' [$\theta$=135°, 315°; Fig. 4(e)]. Additionally, for the pocket 'flat' parts ($\theta$=45°, 225°), the inter-orbital ($d_{xz} \leftrightarrow d_{yz}$), inter-pocket interaction $\Gamma_{xz,yz}$ drives a sign-changing pairing gap [Fig. 4(e)]. In order to explain two pairs of coherence peaks [as seen in the tunneling spectra, Fig. 1(c)], two maxima of the order parameter are needed, which can be achieved with the mentioned pairing interactions being comparable in strength. In a simple approach within modified spin-fluctuation pairing, the second gap maximum, i.e. $\Delta_1$ at $\theta$=45 °/225 °, is obtained from selectively suppressing the coherence of the $d_{xy}$ orbital, while leaving the $d_{xz}$ and $d_{yz}$ more coherent and degenerate. This can be parametrized via orbital-dependent quasiparticle weights $Z_a$, i.e. $\sqrt{Z_{xy}}<\sqrt{Z_{xz}}=\sqrt{Z_{yz}}$ (Supplementary Section VII) (*30*). To summarize, the coherence of $d_{xy}$ orbital is selectively suppressed for relatively 'strengthening' the inter-orbital ($d_{xz}\leftrightarrow d_{yz}$) pairing, such that the weak inter-orbital interaction can still drive a second gap maximum. Meanwhile, the $d_{xy}$ coherence should not be over-suppressed to ensure the intra-orbital ($d_{xy}\leftrightarrow d_{xy}$) pairing remains sufficient for preserving the first gap maximum at $\theta$=135 °/315 °. As shown in Fig. 4(f), the gap structures with two maxima as detected by spectroscopic techniques are well theoretically reproduced when evaluated on the unhybridized Fermi pockets, directly demonstrating the $d_{xz}+d_{yz}$-orbital selectivity for Cooper pairing (for detailed discussions, see Supplementary Section VIII).

The concluded orbital-selective pairing and the anisotropic-$\Delta\sigma_{b.k.}(\theta)$-implied AFSF [Fig. 3(j)] are mutually consistent. $\Delta\sigma_{b.k.}(\theta)$-described AFSF are derived from the local, instantaneous Coulomb repulsions, which are normally larger for intraorbital pairing (*41*). The orbital-dependent Coulomb repulsions result in the orbital-selective pairing interactions (i.e., AFSF) peaked at $Q \approx <2\pi,0>$ for 1-UC FeSe, and then the orbital-selective pairing. Such orbital sensitivity is a unique feature of AFSF pairing, appearing exclusive from the retarded, attractive, and orbitally undifferentiated phononic pairing (*41*). The observed orbital-selective pairing is thus intrinsic to the AFSF for multiorbital superconductivity, and is inversely further substantiated by the detected electron–AFSF coupling identified as the bosonic-kink origin.

Figure 4(g) summarizes the angular dependences of $\Delta_{avg.}(\theta)$, $\Delta\sigma_{b.k.}(\theta)$, and orbital-weight distribution of $d_{xy}$ and $d_{xz}+d_{yz}$ orbitals from our tight-binding model over outer hybridized pockets. Comparison of $\Delta_{avg.}(\theta)$ and $\Delta\sigma_{b.k.}(\theta)$ reveals that they are out-of-phase with a 45 ° phase shift. Note that according to the above quasiparticle-weight-dressed spin-fluctuation theory, the $d_{xz}/d_{yz}$ orbitals are more coherent ($\sqrt{Z_{xz/yz}} > \sqrt{Z_{xy}}$). Evidently, the 90 °×*N*-peaked $\Delta\sigma_{b.k.}(\theta)$ corresponding to the $Q \approx <2\pi,0>$ AFSF coupling is contributed by $d_{xz}/d_{yz}$ orbitals. For the $d_{xz}/d_{yz}$ components with higher coherence, the bosonic coupling is then expected to be stronger, in agreement with the measured self-energy effect $\Delta\sigma_{b.k.}(\theta)$ showing maxima where the $d_{xz}/d_{yz}$ orbital content is largest.

## DISCUSSION AND PERSPECTIVES

Microscopically, the orbital-selective Cooper pairing originates from the orbital-selective correlations via renormalized quasiparticle weights of different orbitals. Similar to other iron chalcogenides (bulk FeSe (*18*), $A_x\text{Fe}_{2-y}\text{Se}_2$ (*19*), and $\text{Fe}_{1+y}\text{Se}_x\text{Te}_{1-x}$ (*20*)), in 1-UC FeSe, while the weakly correlated $d_{xz}/d_{yz}$ orbitals remain itinerant, the $d_{xy}$ orbital is strongly localized by strong correlations (*21*). The degree of electron correlations is reflected in quasiparticle weight $Z$, which, with increasing correlations, decreases from unity in noninteracting systems to zero in Mott insulators (*5*). Accordingly, for 1-UC FeSe, compared with the strongly correlated $d_{xy}$ orbital, the less correlated $d_{xz}/d_{yz}$ orbitals show considerably larger quasiparticle weight as we adopted in the dressed spin-fluctuation theory. In the scattering process of a Cooper pair within the AFSF scenario, the scattering strength is cooperatively determined by a) the quasiparticle coherence of initial and final states, and b) the spin susceptibility $\chi(q)$ (*30, 42*). In consequence, compared with the small-$Z$ $d_{xy}$ orbital, the large-$Z$ $d_{xz}/d_{yz}$ orbitals prevail with significantly lager quasiparticle coherence and spin susceptibility. Therefore, in 1-UC FeSe, the orbital-selective Cooper pairing emerges by suppressing the pair scatterings involving the less coherent $d_{xy}$ orbitals. Within this picture, the observation that the AFSF dominate at $Q \approx <2\pi,0>$ may naturally arise from the orbitally resolved spin susceptibility of the more coherent $d_{xz}/d_{yz}$ orbitals. The spin susceptibility of the less coherent $d_{xy}$ orbital is in such picture highly suppressed via renormalization by the relatively small quasiparticle weight, making negligible contribution to the detected AFSF.

Our results for the angle dependence of pairing strength support the picture of 1-UC FeSe as a high-$T_c$ superconductor with orbital-selective pairing (*10*) in the 2D limit despite quantum fluctuations. We presented an analysis of self-energy deduced from our BQPI data that independently confirm the conclusion. The influence of speculated pseudogap behavior at 60–70 K (*43*) on selective pairing is negligible at 4.3 K, where 1-UC FeSe/SrTiO$_3$ is in deeply Cooper-paired SC regime. Given that FeSe is the building block of iron chalocogenides, the verification of selective pairing in 1-UC FeSe likely indicates its general existence in extensive multiorbital iron-based superconductors, regardless of the SC spectra being different ('*U*'-shaped vs. '*V*'-shaped). The details of SC order parameter (yielding fully gapped or nodal/quasinodal structure) depend on whether the Fermi pockets intersect with the nodal lines of the gap function. Hence, for multiorbital SC systems with distinct Fermi-surface topology (e.g. 1-UC FeSe vs. bulk FeSe), the SC-spectrum lineshapes can be very different even if the underlying pairing mechanism are identical (i.e. both mediated with spin fluctuations, and showing orbital-selective character).

Most fundamentally, the orbital-selective correlations derived from Hund's metallic state are expected to dominate such orbitally selective superconductivity. This suggests even when multiorbitals are involved, only specific orbital(s) is essentially responsible for the pairing as in single-orbital-dominated cuprates. Our finding thus implies the doped 'parent' Mott insulator scenario for driving superconductivity is generally applicable for both copper- and iron-based superconductors. The revealed universal importance of electron correlations in these two main categories of high-$T_c$ superconductors can be of basic significance for a unified formulism of high-$T_c$ SC mechanism.

Atomically manipulating different degrees of freedom and separately revealing their roles in determining various electronic properties are at the heart of modern quantum technologies and future functional quantum-device applications. The discovered orbital-selective pairing uncovers the physical mechanism for how orbital degree of freedom 'shapes' the high-$T_c$ superconductivity microscopically. Explorations by taking into account additional electronic degrees of freedom (charge, orbital, spin, valley) may lead via analogous mechanisms to previously unexpected emergent phenomena. 1-UC FeSe with $d_{xz}/d_{yz}$ orbitals primarily responsible for pairing, in contrast to bulk FeSe dominated by $d_{yz}$ orbital instead (*10*), points to a method for orbital control via tailoring thickness. By further revealing the correspondence between orbitals and more electronic properties, the orbital manipulation with atomic precision may be stimulated for future studies with prospects for detecting orbit-related quantum effects and developing functional orbitronics.

**SUPPLEMENTARY MATERIALS**

SI. STM Characterizations of 1-UC FeSe/SrTiO$_3$

SII. Dynes-Function Fittings

SIII. Empirical $T_c$ Extrapolations

SIV. BQPI Data

SV. Atomically Decorated Tips with Different ***q*** Sensitivity

SVI. Further Discussions Related to Measured Self-Energy Effect

SVII. Calculated Spectral-Weight Distribution of Fe 3*d* Orbitals and SC-Gap Structures

SVIII. Detailed Discussions About Orbital-Selective Pairing Dominated by $d_{xz}+d_{yz}$ Orbitals

Figures S1–S10

Tables S1–S2

References *(44)–(58)*

**Acknowledgements:** The authors acknowledge fruitful discussions with Xiaoqiang Liu, and code assistances from Shusen Ye. **Funding:** This work was financially supported by National Natural Science Foundation of China (No.11888101, and No. 11774008), National Key R&D Program of China (No. 2018YFA0305604 and No. 2017YFA0303302), Beijing Natural Science Foundation (No. Z180010), and Strategic Priority Research Program of Chinese Academy of Sciences (No. XDB28000000). B. M. A. acknowledges support from the Independent Research Fund Denmark (No. 8021-00047B). P. J. H. was supported by the U.S. Department of Energy (No. DE-FG02-05ER46236). **Author contributions:** J.W. conceived and instructed the research. C.L. prepared the samples and carried out the STM experiments. A.K., B.M.A., and P.J.H. provided the theoretical calculations. C.L., A.K., S.Z., and Y.L. analyzed the data. C.L. wrote the manuscript with major revisions from J.W., A.K., B.M.A., and P.J.H. **Competing interests:** The authors declare that they have no competing interests. **Data and materials availability:** All data needed to evaluate the conclusions in the paper are present in the paper and/or the Supplementary Materials. Additional data related to this paper may be requested from the authors.

Supplementary Materials for

# Orbital-Selective High-Temperature Cooper Pairing Developed in the Two-Dimensional Limit

Chaofei Liu, Andreas Kreisel, Shan Zhong, Yu Li, Brian M. Andersen, P. J. Hirschfeld, Jian Wang

**The PDF file includes:**

**SUPPLEMENTARY TEXT**

**I. STM Characterizations of 1-UC FeSe/SrTiO$_3$**

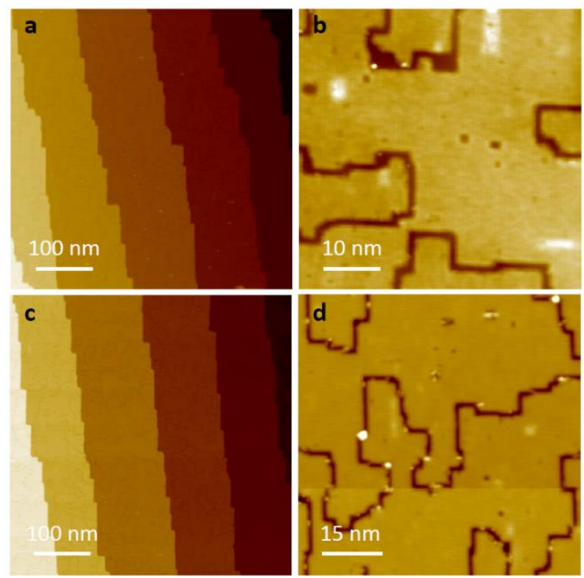

FIG. S1. Topographic images of 1-UC FeSe at different scales, showing (a,c) atomically flat terraces and (b,d) grain boundaries, respectively, demonstrating the high crystalline quality across different spatial scales. Size: **a,c**, 500×500 nm$^2$, **b**, 50×50 nm$^2$, **d**, 70×70 nm$^2$; set point: **a,c**, $V$ = 1 V, $I$ = 500 pA, **b**, $V$ = 0.4 V, $I$ = 500 pA, **d**, $V$ = 0.5 V, $I$ = 500 pA.

## II. Dynes-Function Fittings

### A. Fitting Formulas

Five sets (#1–#5) of temperature-dependent tunneling spectra were acquired. To avoid SC-gap fluctuations over extended spatial regions due to inhomogeneity, each set of spectra at different temperatures were taken at a well-controlled, fixed point within an uncertainty of ~1 nm. The spectra were normalized separately by their respective polynomial backgrounds, which are obtained by the well-established cubic-polynomial fittings to the spectra for bias $|V| \geq 30$ mV (*23, 44*). The normalized (by default) temperature-dependent tunneling spectra were fitted by the Dynes function (*45, 46*)

$$\frac{dI}{dV} = N(E_\mathrm{F}) \int_{-\infty}^{\infty} dE \left[-\frac{\partial f(E+eV)}{\partial eV}\right] \mathrm{Re}\left[\frac{|E-i\Gamma|}{\sqrt{(E-i\Gamma)^2-\Delta^2}}\right],$$

where

$$-\frac{\partial f(E+eV)}{\partial eV} = \frac{1}{k_\mathrm{B}T}\cosh^{-2}\frac{E+eV}{2k_\mathrm{B}T}.$$

Here, $N(E_\mathrm{F})$ is the DOS at $E_\mathrm{F}$, $f$ is the Fermi function, $\Gamma$ is the spectral broadening, and $k_\mathrm{B}$ is the Boltzmann constant. For anisotropic fittings, the Dynes function is alternatively written as

$$\frac{dI}{dV} = \frac{1}{2\pi} N(E_\mathrm{F}) \int_{-\infty}^{\infty} dE \int_0^{2\pi} d\theta \left[-\frac{\partial f(E+V)}{\partial eV}\right] \mathrm{Re}\left[\frac{|E-i\Gamma|}{\sqrt{(E-i\Gamma)^2-\Delta^2(\theta)}}\right].$$

As reported by high-resolution ARPES, $\Delta(\theta)$ on unhybridized electron pocket for 1-UC FeSe shows two different maxima $\Delta_{1,2}$ at $\theta = 45°+180°\times N$ and $135°+180°\times N$ [$N = 0, 1$; $\theta$ defined in Fig. 1(d)] (*24*). Either cos$2\theta$- or cos$4\theta$-type SC-gap function evidently cannot capture the ARPES-measured gap distribution. To incorporate the C2-symmetric gap $\Delta(\theta)$ distribution meanwhile with two gap maxima, the SC-gap anisotropy is introduced by setting

$$\Delta(\theta) = \Delta^{\max}\left[1 - p^\alpha\left(1 - \cos\left[4\left(\theta - \frac{\pi}{4}\right)\right]\right) - p^\beta\left(1 - \cos\left[2\left(\theta - \frac{\pi}{4}\right)\right]\right)\right].$$

The degree of anisotropy is tuned by $p^{\alpha,\beta}$. Two summed cos$4\theta$-type $\Delta(\theta)$ functions with different weights can also be used for fitting the multigap-type SC spectra as previously adopted in (Li$_{1-x}$Fe$_x$)OHFeSe (*28*). However, i) the summed cos$4\theta$-type function would involve more fitting parameters, as compared with the single-cos$4\theta$ case; ii) for 1-UC FeSe, the C4-symmetric cos$4\theta$-type function with only one-valued maxima is inconsistent with the ARPES-measured gap structure.

Experimentally, 1-UC FeSe shows BCS ratios $2\Delta_{1,2}/k_\mathrm{B}T_\mathrm{c}$ of 4.7 and 7.5 for $\Delta_1$ and $\Delta_2$, respectively (*23*), both considerably larger than 3.53 predicted for the weakly coupled BCS superconductor. Influenced by the strong-coupling nature, the tunneling spectrum for 1-UC FeSe is reconstructed by the electron–boson coupling. Accordingly, an additional 'dip–hump' structure is typically induced outside the SC coherence peak (*23*), well beyond the description by Dynes function within the mean-field BCS framework. Therefore, for all Dynes fittings, we selected the bosonic-coupling-unmodified region [−25,25] mV as the fitting window.

The extracted $\Delta_1(T)$ and $\Delta_2(T)$ from the Dynes fittings were fitted by the BCS gap function (*47*)

$$\Delta(T) = \Delta_0 \tanh\left(\frac{\pi}{2}\sqrt{\frac{T_\mathrm{c}}{T} - 1}\right).$$

### B. Anisotropic Dynes Fittings

The five sets of tunneling spectra at different temperatures were at first tentatively fitted by the isotropic Dynes function (results not shown). Evidently, the multigap-type tunneling spectra of 1-UC FeSe cannot be fitted by the isotropic Dynes formula, which only shows single pair of coherence peaks. By adopting two summed Dynes functions with different weights, we found that, while several sets can be constrainedly fitted, the remaining sets show coherence peaks fail to be reproduced by the Dynes formula. At a quantitative level, for all sets of spectra, the obtained SC gaps $\Delta_{1,2}$ (~8.8/~13±0.5 meV), e.g., at 4.3 K, from fittings are strikingly lower than those (~11±0.5/~17±2 meV) directly determined by locating the coherence-peak positions, essentially due to the unreasonably high spectral broadening $\Gamma_{1,2}$.

TABLE. S1. Dynes-fitting parameters [$\Delta_{1,2}(T)$ and $\Gamma(T)$] for the spectra in Fig. 1(c), exemplifying reasonable $\Delta_{1,2}(T)$ with suppressed $\Gamma(T)$. Here, $\Delta_{1,2}$ denote $\Delta(\theta)$ maxima at $\theta = 135°$ and $45°$, respectively (also for Fig. S2 and Table S2).

| $T$ (K) | $\Delta_1$ (meV) | $\Delta_2$ (meV) | $\Gamma$ (meV) |
|---|---|---|---|
| 4.3 | 9.2 | 15.8 | 2.14 |
| 7 | 9.2 | 15.4 | 1.94 |
| 10 | 9.5 | 15.2 | 2.39 |
| 15 | 9.4 | 15.7 | 2.26 |
| 20 | 9.1 | 15.4 | 2.23 |
| 25 | 8.7 | 15.5 | 1.78 |
| 30 | 9.1 | 15.2 | 2.27 |
| 35 | 8.1 | 13.5 | 3.09 |
| 40 | 6.9 | 11.5 | 4.22 |
| 45 | 4.1 | 6.9 | 4.83 |

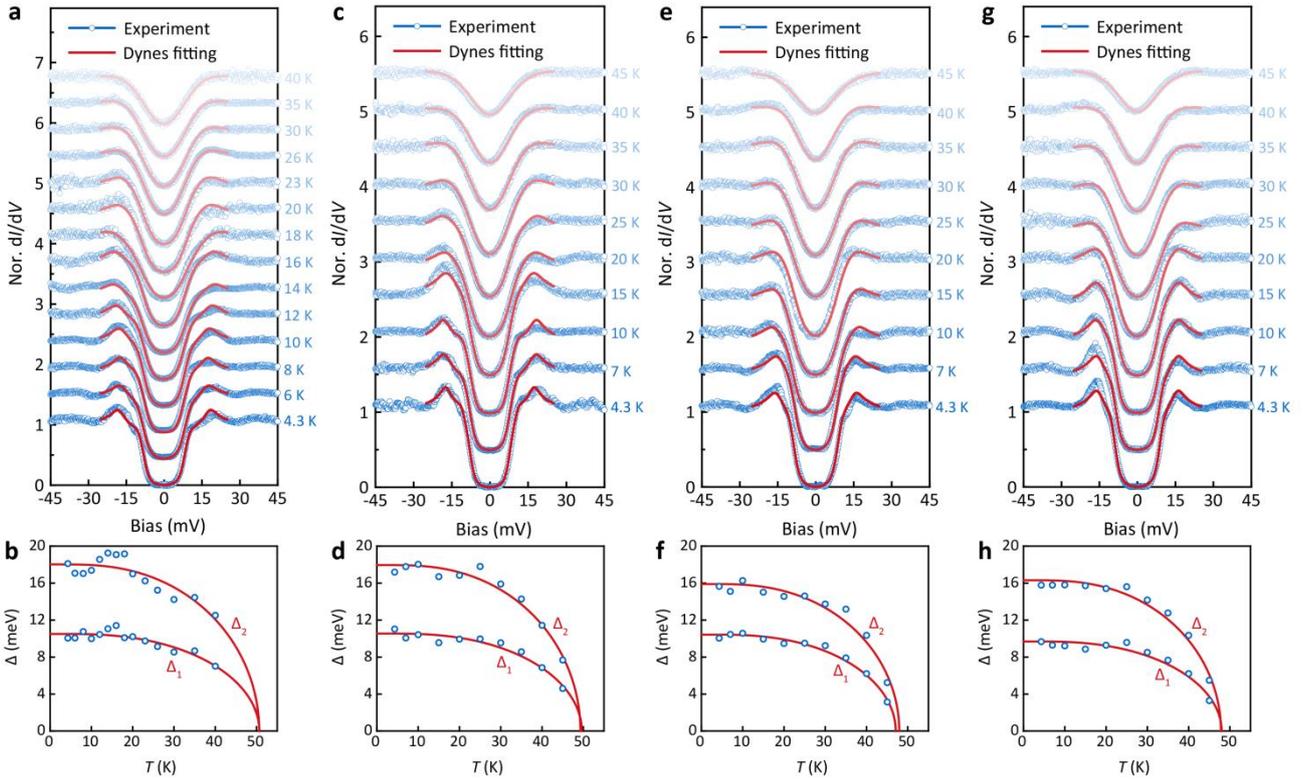

FIG. S2. (a,c,e,g) Anisotropic Dynes fittings (solid curves) to the other four different sets (#1–#4) of temperature-dependent normalized spectra (open symbols) (vertically offset for clarity). (b,d,f,h) BCS fittings (solid curves) to $\Delta_1(T)$ and $\Delta_2(T)$ (open symbols) obtained from the Dynes fittings in (a,c,e,g).

The failure of isotropic Dynes fittings motivates the alternative anisotropic-Dynes approach incorporating the SC-gap anisotropy. The anisotropic Dynes-fitting results are shown in Figs. S2 and 1(c), and Tables S1 and S2. As exemplified in Fig. 1(c) for set #5, the anisotropic fittings well capture the experimental lineshapes, including the coherence peaks. Especially, the yielded $\Delta_{1,2}$ here are comparable with the coherence-peak energies due to low $\Gamma_{1,2}$ (Table S1). These desired anisotropic fittings likewise with reasonable parameters are reproducible for all the other four sets of spectra (#1–#4; Fig. S2 and Table S2). Within the anisotropic-fitting framework, the two pairs of coherence peaks in the tunneling spectra originate from the two different gap maxima [$\Delta_{1,2}$; see, e.g., Fig. 1(d)] on the individual unhybridized electron pocket (*24*). Furthermore, $2\Delta_{1,2}/k_BT_{c1,2}$ calculated from the anisotropic-fitting parameters are ~5 and ~8 for $\Delta_1$ and $\Delta_2$ (Table S2), respectively, signifying the definitive strong-coupling nature aforementioned for 1-UC FeSe.

TABLE. S2. Dynes and BCS fitting parameters [$p^{\alpha,\beta}$(4.3 K), $\Delta_{1,2}$(0 K) and $T_{c1,2}$] for the spectra in Figs. S2 and 1(c).

|    | $p^\alpha$ | $p^\beta$ | $\Delta_1$ (meV) | $T_{c1}$ (K) | $\frac{2\Delta_1}{k_BT_{c1}}$ | $\Delta_2$ (meV) | $T_{c2}$ (K) | $\frac{2\Delta_2}{k_BT_{c2}}$ |
|---|---|---|---|---|---|---|---|---|
| #1 | 0.151 | 0.228 | 10.5 | 50.7 | 4.8 | 18.0 | 50.6 | 8.3 |
| #2 | 0.184 | 0.179 | 10.5 | 49.6 | 4.9 | 18.0 | 49.3 | 8.5 |
| #3 | 0.125 | 0.179 | 10.4 | 47.1 | 5.1 | 15.9 | 47.9 | 7.7 |
| #4 | 0.100 | 0.194 | 9.7 | 48.0 | 4.7 | 16.3 | 47.8 | 7.9 |
| #5 | 0.075 | 0.208 | 9.6 | 50.4 | 4.4 | 16.2 | 50.3 | 7.5 |

### III. Empirical $T_c$ Extrapolations

Besides the BCS fitting to $\Delta_{1,2}(T)$ [e.g., Fig. 1(e)], empirically, $T_c$ can be also determined by extrapolating the temperature-dependent zero-bias conductance (ZBC) for the temperature region near $T_c$. From the five sets of measured normalized spectra, ZBC($T$) were extracted, and linearly extrapolated towards ZBC = 1 separately, where $T = T_c$ (Fig. S3). The extrapolations yield $T_c$ appearing ~10–15 K higher than those by the BCS fittings (Table S4).

Pseudogap exists in the region where Cooper pairs preform, but without long-range coherence. In experiments, the concrete signal for the existence of pseudogap states is normally concluded from the difference between $T_c^{transport}$ and $T_g$, specifying that the pseudogap exists in a temperature region of [$T_c^{transport}$,$T_g$]. (Here, $T_g$ denotes the gap-closing temperature.) Given the increasing curvature of ZBC($T$) at higher temperatures, the linear extrapolation only works well near $T_c$ and otherwise tends to overestimate $T_c$. For example, for set #1 in Fig. S3(a), the linearly extrapolated $T_c$ even exceeds 100 K. $T_c^{BCS}$ and $T_c^{ZBC}$ are unlike $T_c^{transport}$ and $T_g$. ($T_c^{BCS}$ and $T_c^{ZBC}$ denote $T_c$ obtained by extrapolations from BCS fits and temperature-dependent ZBC, respectively.) Thus, the difference between $T_c^{BCS}$ and $T_c^{ZBC}$ may just arise from the different errors in BCS fittings for unconventional superconductors and ZBC extrapolations from low-temperature results.

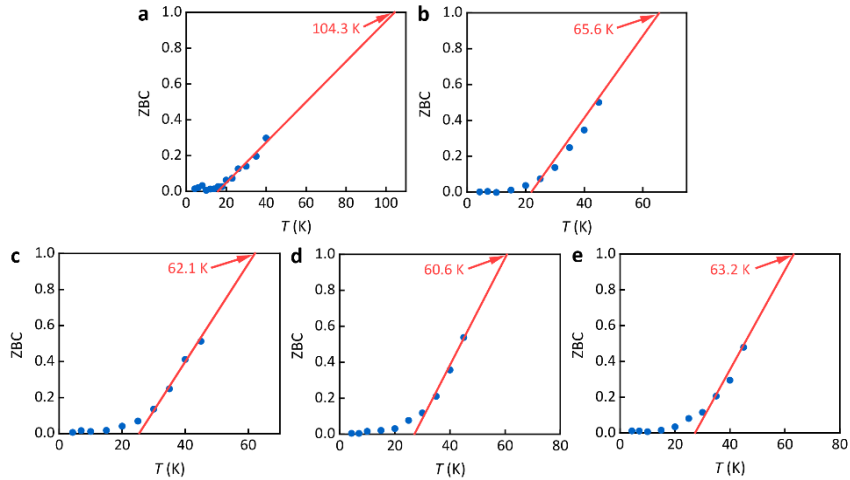

FIG. S3. Temperature-dependent ZBC (solid symbols) extracted from the five sets of normalized experimental spectra. Empirically linear-extrapolated $T_c$ at ZBC = 1 are indicated.

### IV. BQPI Data

## A. Drift Correction by Lawler–Fujita Algorithm

The thermal effect and piezoelectric hysteresis in STM experiments inevitably induce tip drift and, thus, local, spatially evolving distortions during the spectroscopic imaging. To raise the signal-to-noise ratio in the d$I$/d$V$($r$,$E$) mappings and accordingly improve the quality of related FFT-BQPI patterns, the distortions inferred from the topographic images were corrected in the simultaneously acquired d$I$/d$V$($r$,$E$) mappings by Lawler–Fujita algorithm (48). As exemplified in Fig. S4, a typical topographic image and the corresponding FFT pattern before and after applying the Lawler–Fujita drift-correction algorithm are presented for comparison. Compared with the original FFT image [Fig. S4(c)], the drift-corrected FFT pattern [Fig. S4(d)]] shows that the pixels of the Bragg point are moderately reduced. In the limit of perfectly registered lattice without drift, the Bragg point is expected to be ideally peaked in intensity and only consists of a single pixel. The reduced Bragg pixel by the algorithm here suggests the improved topographic quality by largely excluding the extrinsic drift influence.

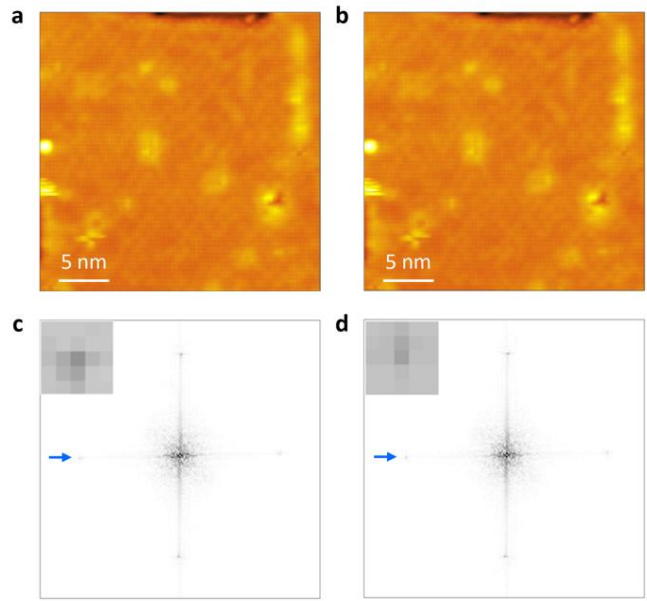

FIG. S4. Topographic image and corresponding FFT pattern (a,c) before and (b,d) after the drift correction. The insets in (c) and (d) are the zoom-in views of the Bragg points (arrows). (a,b) 28×28 nm$^2$; set point: $V$ = 0.1 V, $I$ = 500 pA.

## B. FFT-BQPI-Pattern Processing Steps

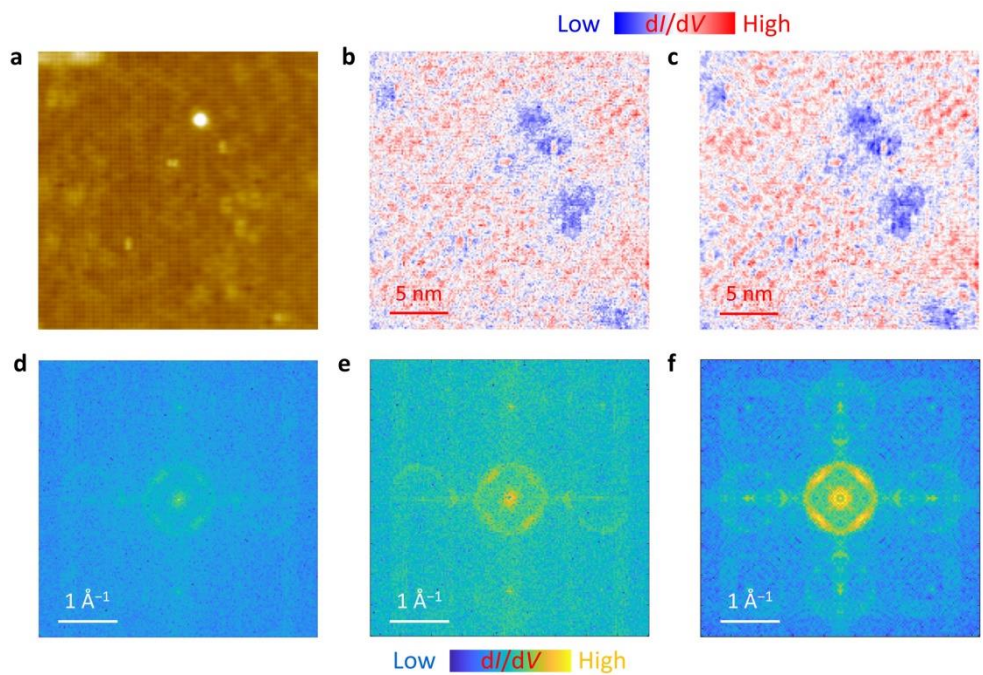

FIG. S5. Exemplifying the FFT-BQPI-pattern processing procedure. (a) Topographic image (24×24 nm$^2$; set point: $V$ = 0.04 V, $I$ = 2500 pA). (b,c) Simultaneously acquired BQPI d$I$/d$V$($r$,$E$) mapping at 23 meV (b) before and (c) after drift correction [adapted as Fig. 2(a)]. (d) Raw FFT pattern of the d$I$/d$V$($r$,$E$) mapping in (c). (e) Gaussian-filtered FFT pattern. (f) Gaussian-filtered and symmetrized FFT pattern [adapted as Fig. 2(b)].

Mainly three steps were successively used to optimize the signal-to-noise ratio in our FFT-BQPI patterns (e.g., Fig. S5). i) First, the BQPI d$I$/d$V$($r$,$E$) mappings were drift-corrected by the Lawler–Fujita algorithm [Fig. S5(c)] (48). ii) Second, for clarity of interested scattering features, the high-intensity origins $q$ = (0,0) of the FFT-BQPI patterns were suppressed by the Gaussian function with kernel at $q$ = (0,0) [Fig. S5(e)]: FFT-BQPI$_{Gausian}$ = FFT-BQPI$_{non-Gaussian}$[1−Gaussian($q$ = (0,0), $\sigma$)] (49). The central peaks at $q$ = (0,0) essentially stem from the randomly scattering defects and the long-range variations of the surface. The Gaussian filtering of these origin peaks in FFT-BQPI patterns does not affect the scattering signals of interest. iii) Third, the FFT-BQPI patterns were four-fold symmetrized [Fig. S5(f)] given the C4 symmetry of the Fermi surface.

## C. Selected BQPI Patterns

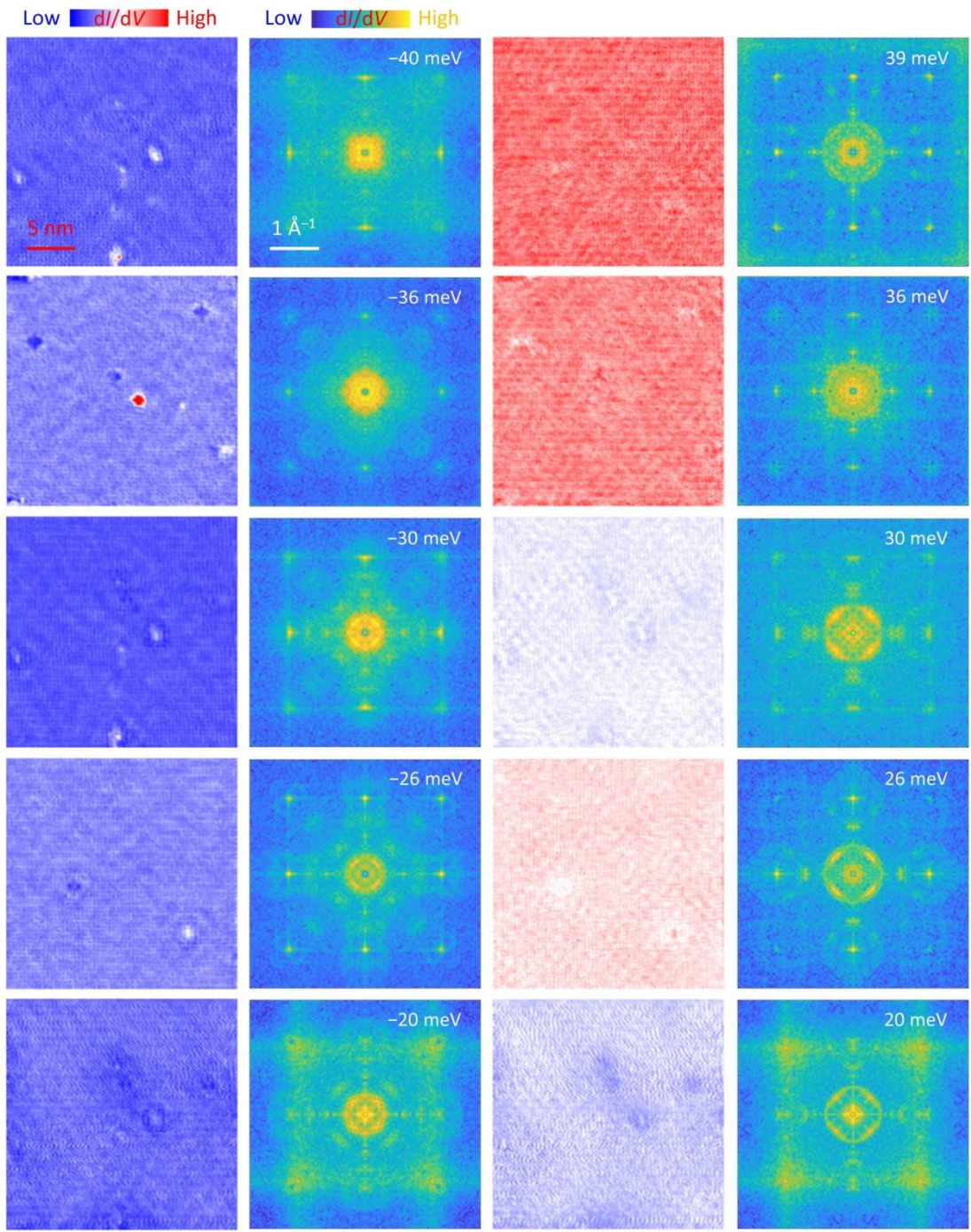

FIG. S6. Selected BQPI d$I$/d$V$($r$,$E$) mappings (24×24 nm$^2$) and corresponding FFT patterns, separately sharing the respective color-scale bars above the panels.

## V. Atomically Decorated Tips with Different $q$ Sensitivity

By frequently preparing new PtIr tips and constantly applying voltage pulse ($V$ = 0.1–1 V, $\Delta t$ = 10 ms), we realized a relatively high resolution selectively for low-$q$ scatterings (Fig. S7), which likely arises because of the change of atomic configuration decorated at the tip end at a microscopic scale (*10, 26*). In general situations, it is difficult to resolve the small- and large-wavelength quasiparticle scatterings simultaneously. The STM tips with different atomic 'sharpness' selectively detect with relatively high resolution the BQPI modulations with different wavelengths (*10, 26*). Specifically, the high-spatial-resolution tip (tip α) is more sensitive to the short-wavelength BQPI modulations, which occur in high-$q$, inter-band scatterings among strikingly different parts of the BZ, and vice versa (tip β). In our experiments, for FFT-BQPI pattern before intense tip treatments [e.g. Fig. S7(a)], the inter-pocket quasiparticle scatterings $q_2$ and $q_3$ (solid arrows) are selectively resolved with relatively higher resolution. Meanwhile, the intra-pocket quasiparticle scattering $q_1$ (dashed arrows) remains resolved, but is dominated by low-$q$ mussy scatterings showing no fine structure. Inversely, for FFT-BQPI pattern after intense tip treatments [e.g. Fig. S7(b)], the intra-pocket quasiparticle scattering $q_1$ (solid arrows) is instead selectively high-resolved with remarkable intensity anisotropy. However, the inter-pocket quasiparticle scatterings $q_2$ and $q_3$ (dashed arrows) are weakly or even barely resolved. These FFT-BQPI data with different $q$ resolution due to intense tip-treatment procedures can be most likely explained by the atomic-scale change of decorated tip shapes. In such a scenario, Fig. S7(a) and Fig. S7(b) can be imaged by the tips α and β, respectively.

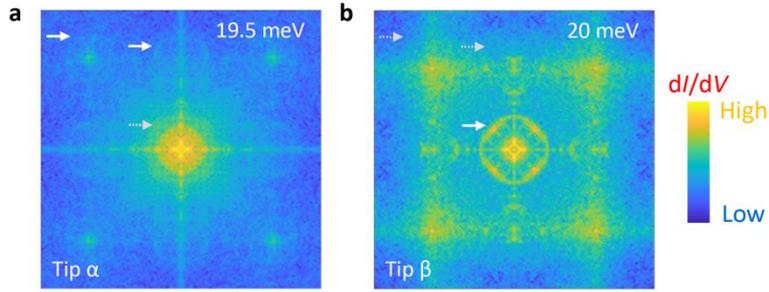

FIG. S7. Typical FFT-BQPI patterns before and after intense tip treatments with large- and small-$q$ (adapted from Fig. S6) resolution, respectively, which may be imaged by the STM tips with different $q$ sensitivity.

Due to the electron-type nature of $M$-centered Fermi pockets, the $q_1$ pocket at negative binding energies is increasingly small in size as departing from $E_F$ [Fig. 2(e)]. In this case, the scattering-intensity anisotropy is beyond the spectral resolution and will not be further analyzed.

## VI. Further Discussions Related to Measured Self-Energy Effect

### A. ReΣ Definition by Difference Between Measured $E(q)$ and Noninteracting Backgrounds

Since describing the deviation from free-electron dispersion, the real part of self-energy ReΣ can be alternatively defined as the difference between measured $E(q)$ and noninteracting backgrounds [Fig. S8(a)]. Given that obtaining high energy resolution in FFT-BQPI patterns is quite challenging, the ReΣ precisely defined by this conceptual method may be not quantitatively accurate, but can serve as an independent check of its angular dependence. Consistently, as Fig. 3(h), the extracted ReΣ($E$) at selected directions also shows bosonic-mode-reconstructed peaks at $E^\Sigma$ ~37 meV [Fig. S8(a)]. Figure S8(b) plots ReΣ($E^\Sigma$ ~37 meV) vs. $\theta$, revealing similar highly anisotropic feature as $\Delta\sigma_{b,k}(\theta)$ in Fig. 3(j).

Note that the extracted ReΣ($E$) in Fig. S8(a) and $E(q)$ in Fig. 3(h) show gentle kink-like features below 30 meV, especially for $\theta$ = 11 ° and 22 °. However, these kinks are rather weak compared with those near $E^\Sigma$ ~37 meV, and are obviously beyond the resolution of current BQPI technique for a convincing analysis. For most ReΣ($E$) and $E(q)$ curves, only a single kink was detected at a common energy of 35–40 meV within resolution, probably implying the role of Hund's rule coupling in aligning the bosonic-kink energy scales $E^\Sigma_\gamma$ for different orbitals $\gamma$ (*50*).

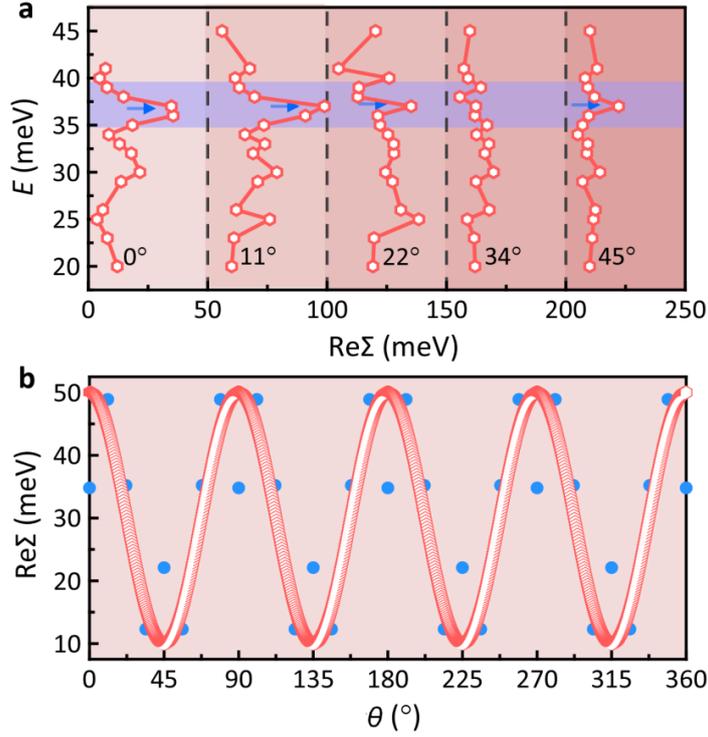

FIG. S8. (a) ReΣ($E$) along representative directions (horizontally offset for clarity), which was determined from the departure of measured $E(\boldsymbol{q})$ from the noninteracting backgrounds [Fig. 3(h)]. The arrows highlight ReΣ maxima at $E^\Sigma$ ~37 meV. (b) Measured ReΣ(37 meV) as a function of $\theta$.

### B. Intrinsicity of Fluctuations in $\Delta\sigma_{b.k.}(\theta)$

Based on the following considerations, the measured angular-dependent self-energy effect $\Delta\sigma_{b.k.}(\theta)$ [Fig. 3(j)] is unlikely attributable to the fluctuating errors in experiments. i) The dispersion kinks used for extracting $\Delta\sigma_{b.k.}(\theta)$ coincide in energy scale with the spectral hump reconstructed by electron–boson interaction [Fig. 3(h) vs. Fig. 3(i)] (*23*); such energy consistency suggests an intrinsic bosonic-coupling origin of the detected kink features. ii) The $\Delta\sigma_{b.k.}(\theta)$ peaks at $\theta = 0°$, 90°, 180°, and 270°, corresponding to the directions of inter-pocket nesting vectors $\boldsymbol{Q} \approx \langle 2\pi,0 \rangle$ for AFSF within extended $s_\pm$- and nodeless $d$-wave pairing scenarios (Fig. S9). For decisive estimation on $\Delta\sigma_{b.k.}(\theta)$ fluctuations, repeating our experiments is recommended in future studies by using even higher-resolution BQPI, e.g., at a base temperature far below 1 K by a $^3$He or a dilution refrigerator. Further combined with the elaborate STM tip, the highly challenging self-energy effects with band-resolution (*51*) may be realized, which would put substantial constraints for quantitative theoretical study of the Cooper-pairing mechanism.

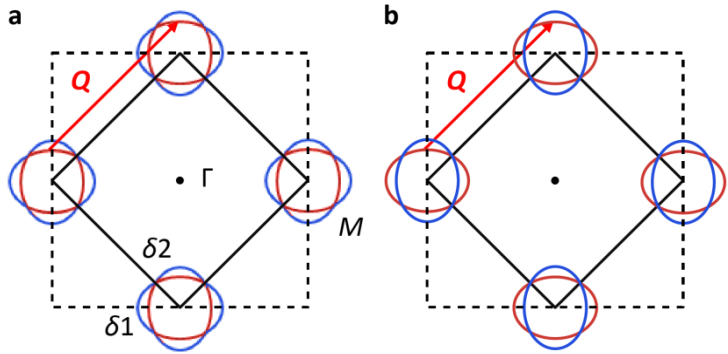

FIG. S9. Fermi-surface topologies of 1-UC FeSe/SrTiO$_3$ in the folded BZs (solid lines) for (a) extended $s_\pm$- and (b) nodeless $d$-wave pairings. $\boldsymbol{Q} \approx \langle 2\pi,0 \rangle$ schematically denotes the inter-pocket pair-scattering vectors.

### C. Electron–AFSF Coupling in Cooperative-Pairing Picture

The electron–AFSF coupling concluded responsible for the anisotropic self-energy reconciles previous observations of gap minima at intersections of hybridized ellipse-like electron pockets (*24*), magnetic-excitation-like bosonic mode (*23*), and sign reversal manifested by nonmagnetic-scattering-induced bound states (*52*) in 1-UC FeSe. Theoretically, the heavily electron-doped 1-UC FeSe with checkerboard (likely) antiferromagnetism (*53, 54*) can be described by Anderson-lattice model for localized spins coexisting with itinerant electrons (*37*), despite the debate over concrete AFM spin order (collinear-, block-checkerboard-, pair-checkerboard-, and checkerboard-type, etc.). In one popular scenario, Cooper pairing arises predominantly from exchange of spin fluctuations, but can be assisted by interfacial forward scattering phonons, leading to a high critical temperature of ~ 60 K (*13*). According to this 'cooperative pairing' picture, the detected AFSF coupling originates from the electronic-pairing channel intrinsic to heavily electron-doped FeSe. The absence of phonon signal in STM experiments can be interpreted by the following facts. i) The interpocket AFSF pair scatterings dominate over the intrapocket phononic pair scatterings (*37, 55*). ii) STM is a local, surface-sensitive technique, incapable of capturing the interface phononic coupling beneath the FeSe film.

## VII. Calculated Spectral-Weight Distribution of Fe 3*d* Orbitals and SC-Gap Structures

The electronic-structure calculation for 1-UC FeSe is based on a tight-binding model for bulk FeSe (at high temperatures) (for details, see Ref. (*30*)). The 1-Fe unit cell that corresponds to the unfolded BZ was adopted. Since the strong electron doping of 1-UC FeSe by $SrTiO_3$ substrate suppresses the spin-density-wave order (*56*), we started with a paramagnetic DFT band structure of the usual type. Several modifications were taken into account for bulk FeSe to 'simulate' 1-UC FeSe: i) all out-of-plane hoppings were ignored to guarantee a strictly 2D character; ii) orbital order was neglected since it is never observed in 1-UC FeSe; iii) FeSe system should be electron-doped. By DFT calculations, $SrTiO_3$ was previously shown only serving to electron-dope the free-standing FeSe film (*53*). We thus considered the substrate effect on the band structures of 1-UC FeSe by shifting $E_F$ to match the size of Fermi pockets as measured in ARPES (*24*). Furthermore, the BZ folding was considered within the (quasi-) nodeless *d*-wave pairing scenario.

For the gap-structure calculation, the standard spin-fluctuation pairing theory was used, but with modified quasiparticle weights $Z$ for different orbitals, where the coherence of $d_{xy}$ orbital is selectively suppressed ($\sqrt{Z_{xy}}$=0.4273, $\sqrt{Z_{xz}}$=0.9826, $\sqrt{Z_{yz}}$=0.9826). Technically, such orbital selectivity is incorporated by adopting that: i) $c_l^\dagger(\bm{k})$ create quasiparticles with weight $\sqrt{Z_l}$ in orbital $l$, $c_l^\dagger(\bm{k}) \rightarrow \sqrt{Z_l} c_l^\dagger(\bm{k})$; ii) the calculation of spin susceptibility includes the renormalized Green's function, $G_{ll'}(\bm{k},\omega_n) \rightarrow \sqrt{Z_l}\sqrt{Z_{l'}} G_{ll'}(\bm{k},\omega_n)$ (for details, see Ref. (*30*)).

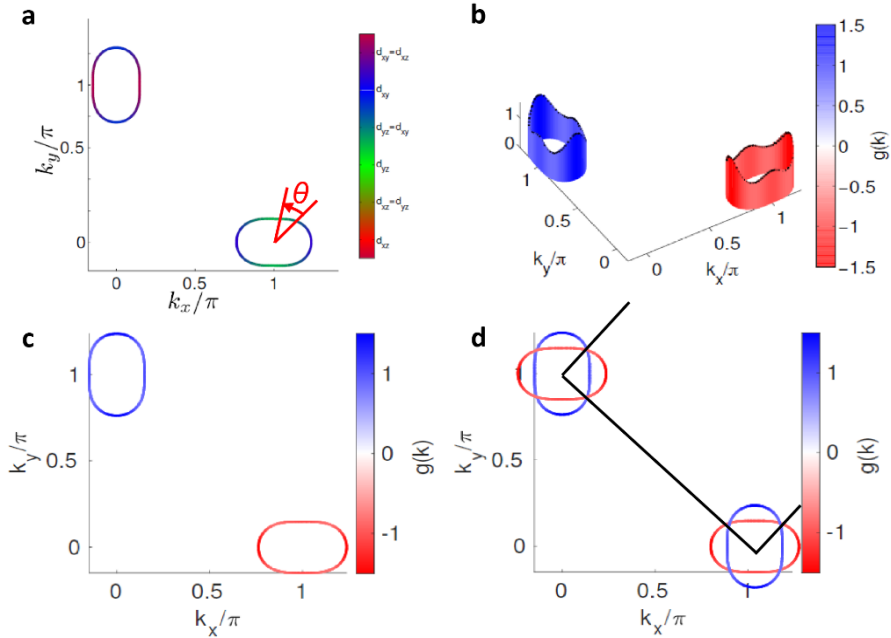

FIG. S10. (a) Fermi surface of our model describing 1-UC FeSe color-encoded with orbital character before BZ folding. (b–d) Calculated gap symmetry function $g(\boldsymbol{k})$ [proportional to the gap below $T_c$, $g(\boldsymbol{k}) \propto \Delta(\boldsymbol{k})$] over the Fermi surface (b,c) before and (d) after BZ folding. The red and blue colors represent the different signs of order parameter. Light gray (black) axes define the coordinate axes in the unfolded (folded) BZ.

Consistent with ARPES experiments (*57*), the calculated Fermi-surface topology consists of only *M*-centered electron pockets (Fig. S10). More accurately, the ellipse-like fine structures of the electron pockets were also well reproduced. The obtained Fe 3*d* orbital-weight distributions and gap structures over the Fermi surface are shown in detail in Fig. S10.

## VIII. Detailed Discussions About Orbital-Selective Pairing Dominated by $d_{xz}+d_{yz}$ Orbitals

In the normal state of iron chalcogenides, the $d_{xz}/d_{yz}$ orbitals are degenerate at *M* points. Even after transition into the SC state, the $d_{xz}/d_{yz}$ degeneracy remains preserved under the protection of robust C4 symmetry, despite an *M*-centered gap between electron and hole bands influenced by the orbital-dependent band renormalization and $d_{xz}/d_{xy}$ band hybridization (*21*). Accordingly, the absence of nematic orbital order in SC 1-UC FeSe requires the quasiparticle weights for $d_{xz}$ and $d_{yz}$ orbitals be degenerate. These less correlated and degenerate $d_{xz}/d_{yz}$ orbitals both correspond to much higher quasiparticle weights, i.e., higher quasiparticle coherence than $d_{xy}$. In contrast, the more correlated $d_{xy}$ orbital possesses narrow electronic bandwidth besides lower quasiparticle coherence. The shrinked electronic bandwidth results in narrow spin-excitation spectra and, thus, weak effective magnetic exchange coupling (*58*), which is unbeneficial for superconductivity within the AFSF-pairing scenario. Thereby, the $d_{xz}/d_{yz}$ orbitals are physically reasonable in cooperating for Cooper pairing.

The $d_{xz}/d_{yz}$-selective pairing is also reconciled within the electron-hopping picture regarding nematic order. By BQPI, bulk FeSe crystals were demonstrated that the Cooper pairing is selectively driven by $d_{yz}$ orbital (*10*). With the low-energy physics dominated by $d_{yz}$ orbital, the electron hopping is preferred along *y* direction over *x* direction. Such highly anisotropic hopping agrees with the nematicity in FeSe. Inversely, in 1-UC FeSe, the $d_{xz}/d_{yz}$-selective pairing implies the electron hoppings show no direction preference, consistent with the absence of nematic phase therein.